\documentclass[dvipdfm]{article}
\usepackage[dvips]{graphicx} 
\usepackage[usenames,dvipsnames]{color}
\usepackage[dvipdfm]{hyperref} 

\newcommand{\bq}{\begin{eqnarray}}
\newcommand{\eq}{\end{eqnarray}}

\begin{document}

\title{\bf Pricing Variable Annuity Contracts with High-Water Mark Feature}
\author{V.M. Belyaev 
\\  \it Allianz Investment Management,
    Allianz Life \\
  \it Minneapolis, MN, USA}

\maketitle
\begin{abstract}Variable annuities (VA) are popular insurance products.
VAs provides the insured with a guaranteed accumulation rate on their premium at maturity.
In addition, the insured may receive extra benefit if returns of underlying funds are high enough.
Here we consider a special case of VA with high-water mark feature and Guaranteed Minimum payment reset.
In Black-Scholes model for underlying fund  we derive explicit pricing formula for this type of contract.
The value of VA  contracts depends on the time between observation dates.
Corrections due to this effect are calculated and compared with Monte-Carlo results.
Good agreement between analytical formula and numerical calculations of VA values is demonstrated.
\end{abstract}

\section{Introduction}
Variable annuity (VA)  contract provides the policyholder with a guaranteed minimum return and offers extra benefit if returns of underlying funds are high enough.
VAs with high water mark feature are similar to lookback option: they choose the maximal fund value over the entire term of an annuity for calculating
payoff at maturity. Within Black-Scholes framework \cite{BS1973}, these  type of options were considered in  \cite{T2000},\cite{GS2003},\cite{Lee2003}.
The effects of stochastic interest rates and mortality were explored in \cite{LT2003}.

In this paper, we focus on valuation of VA contract value where guaranteed minimum payoff is calculated
from set of fund values over the entire term of contract.
High-water mark values are calculated at predetermined observation dates.  Usually, frequencies of  these reset dates are fixed and they can be monthly, quarterly or yearly.
Because of this feature, some of maximal fund values can be missed and it reduces value of the contract.
This effect needs to be taken  into account to price correctly VA contract.

\section{Contract Payoff Definition}We consider the following lookback contract, which pay-off at expiration time $T$ can be presented in the following form
\bq
Payoff=V(T)={\cal N}Max(Hw(T),(1+\gamma\Delta t)V(t-\Delta t))
\label{liab}
\eq
where ${\cal N}$ is a notional \footnote{Below we assume that ${\cal N}=1$ and $S(0)=1$}; $V(0)=1$;
\bq
 Hw(t)=Max(S(t)/S(0),Hw(t-\Delta t));
\nonumber
\\
Hw(0)=1;\;\;
\eq
is a High Watermark; $S(t)$ is price of underlying at time $t$; $\Delta t$ is  time interval between observation dates;
 $\gamma$ is a rate of guaranteed minimum payoff. 

In continuous limit $\Delta t\to 0$ equation (\ref{liab}) can be presented in the following form
\bq
V(T) & = & Max\left[
\frac{S(t)}{S(0)}e^{\gamma(T-t)}|t\in[0,T]
\right]=
\nonumber
\\
& = & e^{\gamma T}Max\left[
\frac{S(t)}{S(0)}e^{-\gamma t}|t\in[0,T]
\right]
\label{liabc}
\eq
From eq.(\ref{liabc}) we can see  that the probability distribution of the VA  contract value is distribution of   extreme (maximal) values of 
underlying
\bq
\tilde S(t)=S(t)e^{-\gamma t}
\label{ts}
\eq
So, we can use distribution of extreme (maximal) values  to calculate present value of this VA contract.

\section{Extreme Value Distribution in Black-Scholes Model}

Consider the following dynamics for underlying $S(t)$
\bq
d\ln(S(t))=\mu dt +vdW(t)
\label{BS}
\eq
where
\bq\mu=r-y-\frac12 v^2
\eq
is a risk-neutral drift; $r$ and $y$ are short term interest rate and dividend yield; $v$ is volatility; $W(t)$ is a standard Brownian motion
\bq
<dW(t_1)dW(t_2)>=\delta(t_1-t_2)dt
\eq
Notice, that to model distribution of underlying $\tilde S(t)$ (\ref{ts}), all we need to do, is just to redefine dividend yield as
\bq
y\to y+\gamma
\eq

Probability distribution function of returns
\bq
x(t)=\ln(S(t)/S(0))
\eq
 in Black-Scholes world (\ref{BS}) satisfies the following Fokker-Plank equation
\bq
\frac{\partial P(x,t)}{\partial t}=\frac{v^2}{2}\frac{\partial^2 P(x,t)}{\partial x^2}-\mu\frac{\partial P(x,t)}{\partial x}
\label{diff}
\eq
with  initial boundary condition
\bq
P(x,0)=\delta(x)
\label{bc}
\eq
This solution is 
\bq
P_0(x,t)=\frac1{\sqrt{2\pi v^2 t}}e^{-\frac{(x-\mu t)^2}{2v^2 t}}
\eq

To find probability distribution function of maximal values $x(t)$ for time interval $0<t<T$ it is convenient to calculate probability of event  that $x(t)<h$ for all $t$ in the time interval $[0,T]$.
This problem is equivalent to problem with absorbing wall at $x=h$ and therefore we need to find solution which satisfies eq.(\ref{diff}) for all $0<t<T$ and $x<h$ 
and to impose the following boundary conditions:
\bq
P(x=h,t)=0.
\label{bc1}
\eq
Solution of eq.(\ref{bc1}) with boundary conditions (\ref{bc1}) can be found from the following anzatz:
\bq
P(x,t,h)=P_0(x,t)+A P_0(x-x_0,t)
\label{anzts}
\eq
where $A$ and $x_0$ are constants and $x_0>h$.

Eqs. (\ref{bc1},\ref{anzts}) lead us to the following equation
\bq
P(h,t,h)=\frac1{\sqrt{2\pi v^2 t}}e^{-\frac{(h-\mu t)^2}{2v^2 t}}\left(1
-Ae^{-\frac{\mu x_0}{v^2}}e^{-\frac{x_0(x_0-2h)}{2v^2t}}
\right)=0.
\eq
and we obtain that
\bq
A=e^{2\frac{\mu h}{v^2}};\;\;  x_0=2h.
\label{pars}
\eq
Then, probability of the event, that return $x$ does not cross the border $x=h$ by time $t=T$ can be calculated 
\bq
& & P(x<h)=\int_{-\infty}^hP(x,h,T)dx=
\nonumber
\\
& & =\frac12\left(1+erf\left(\frac{h-t\mu}{\sqrt{2v^2t}}
\right)-
e^{2\frac{\mu h}{v^2}}
erfc\left(\frac{h+t\mu}{\sqrt{2v^2t}}\right)
\right)
\eq
where
\bq
& & erf(z)=\frac2{\sqrt{\pi}}\int_0^ze^{-x^2}dx
\nonumber
\\
& & erfc(z)=1-erf(z)
\eq
are error functions.

It gives us the following formula for  probability distribution of maximal values
\bq
 & & P(h,\mu,v,T)  =  \frac{\partial}{\partial h}P(x<h)=
\nonumber
\\
& &  =  \frac2{\sqrt{2\pi v^2 T}}
e^{-\frac{(h-T\mu)^2}{2v^2T}}
-
\frac{\mu}{v^2}e^{2\frac{\mu h}{v^2}}
erfc\left(\frac{h+T\mu}{\sqrt{2v^2T}}
\right)
\label{hdistr}
\eq

\section{High Watermark Liability Value}

In the  limit $\Delta t\to 0$ present value of VA contract (\ref{liabc}) can be calculated from maximal value distribution (\ref{hdistr}) as
\bq
V(S,X,T)= e^{(\gamma-r) T}Max\left[
S(t)e^{-\gamma t}|t\in[0,T],X
\right]=
\nonumber
\\
= e^{(\gamma-r) T}X+e^{(\gamma-r) T}\int_{\ln(X/S)}^\infty (Se^h-X)P(h,\mu,v,T)dh=
\nonumber
\\
=e^{\gamma T}\left(X e^{-rt}+2\;Call(S,X,v,r,y+\gamma,t)+
\right.
\nonumber
\\
+e^{-rt}\left\{
\frac12X
\left(
1+erf\left(\frac{\mu t-\ln(X/S)}{\sqrt{2v^2t}}
\right)
\right)-
\right.
\nonumber
\\
\left.
-\frac{S}{v^2+2\mu}\left[
\mu\left(
1+erf\left(
\frac{t(v^2+\mu)-\ln(X/S)}{\sqrt{2v^2t}}
\right)\right)e^{(\mu+v^2/2)t}+
\right.\right.
\nonumber
\\
\left.\left.\left.
+\frac12\left(
\frac{X}{S}
\right)^{1+\frac{2\mu}{v^2}}
v^2 erfc\left(
\frac{\mu t+\ln(X/S)}{\sqrt{2v^2t}}
\right)
\right]
\right\}\right);
\label{liabh}
\eq
 where  $\mu=r-y-\gamma$ ;  $X=S_H\geq S(0)$ is a strike;  $S_H$ is  maximal observed price  of underlying from issue date to current  time ($t=0$).  $Call(S,X,v,r,y,t)$ is a Black-Scholes formula for  European Call Option price. 
Here, to calculate present value of that contract we  take into account discount factor $e^{-rT}$.

 High Watermark contract value depends on size of time interval between observation dates ($\Delta t>0$).
Value of contract with $\Delta t>0$  is lower
than  in the limit $\Delta t\to 0$
 just because selected dates of resets (observation dates)  can miss the actual extremal values of underlying stochastic process.

To take into account finite size of time steps, let us consider Brownian motion (\ref{BS}) on time interval $T$ with $N$ equal time intervals between observation dates:
\bq
0, t_1, t_2,\dots ,t_n,\dots ,T;
\nonumber
\\
t_n=n\;dt;\;\; dt=T/N.
\label{list}
\eq

Then, the average maximal value near the point $t_i$ is
\bq
& & \left<
 Max(e^{x(t)}|t\in [t_i-\Delta t/2,t_i+\Delta t/2])
\right>_h=
\nonumber
\\
& & =e^{x(t_i)}\left(1+v\sqrt{2\frac{dt}{\pi}}+O(dt)\right)
\eq
It means that in the limit of small time steps $\Delta t\to 0$, distributions of extreme values can be presented in the following form
\bq
P(h,\mu,v,T,\Delta t)\simeq P(\tilde h,\mu,v.T,\Delta t=0)
\eq
where 
\bq
\tilde h\simeq h+v\sqrt{2\frac{dt}{\pi}}.
\eq
Then, the  integral in (\ref{liabh}) can be transformed to the following form
\bq
& & \int_{\ln(X/S)}^\infty (Se^h-X)P(h,\mu,v,T,\Delta t)dh\simeq
\nonumber
\\
& & \simeq\int_{\ln(X/S)}^\infty (Se^h-X)P(h+\epsilon,\mu,v,T,\Delta t=0)dh=
\nonumber
\\
& & =
 e^{-\epsilon}\int_{\ln(\tilde X/S)}^\infty (Se^{h}-\tilde X)P(h,\mu,v,T,\Delta t=0)dh
\label{corr}
\eq
where $\epsilon=v\sqrt{2\frac{\Delta t}{\pi}}$; $\tilde X=Xe^\epsilon$.

And finally, from  (\ref{liabh}) we obtain the following formula for VA contract value 
\bq
& & Xe^{(\gamma-r)T}+e^{-rT}\int_{\ln(X/S)}^\infty (Se^h-X)P(h,\mu,v,T,\Delta t)dh\simeq
\nonumber
\\
& &  \simeq  Xe^{(\gamma-r)T}+e^{-\epsilon} \int_{\ln(\tilde X/S)}^\infty (Se^{h}-\tilde X)P(h,\mu,v,T,\Delta t=0)dh
\eq
where $X=S_He^{\gamma t_h}$; $t_h$ is a time from the observed underlying value to current time which gives the highest
guaranteed payoff at time $t=0$.

\begin{figure}
\begin{center}
\includegraphics[width=\textwidth, bb=0 0 500 300]{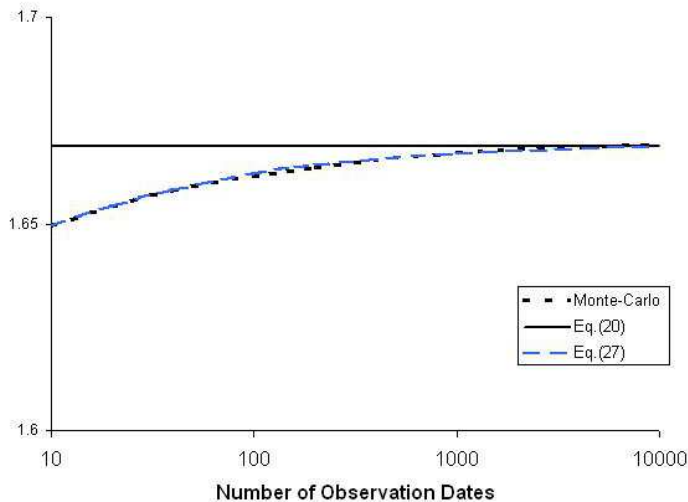}
\end{center}
  \caption{VA Contract Value vs Number of Observation Dates.
$T=10$, $r=5\%$, $\gamma=8\%$, $v=10\%$ 
}
 \label{fig_LibH}
\end{figure}

At issue date, the initial contract value is
\bq
& & Se^{(\gamma-r)T}+e^{-rT}S\int_{0)}^\infty (e^h-1)P(h,\mu,v,T,\Delta t)dh\simeq
\nonumber
\\
& &  \simeq  Se^{(\gamma-r)T}+e^{-\epsilon}S \int_{0}^\infty (e^{h}-1)P(h,\mu,v,T,\Delta t=0)dh
\label{initv}
\eq
Fig.(\ref{fig_LibH}) demostrates a good agreement between analytical formula (\ref{initv}) and numerical calculations of VA values.

\section{Conclusion}
In this paper we consider VA contract which guaranteed minimum payoff is calculated
from a set of maximal fund values over the entire term of contract.
In the limit of small time intervals between observation dates ($\Delta t\to 0$) we derived explicit pricing formula for this type of VA contracts.
Finite time interval corrections are calculated and compared with Monte-Carlo results.
Good agreement between analytical formula and numerical calculations of VA values is demonstrated.

\end{document}